\documentclass[aps,prl,epsfig,showpacs,floats,twocolumn,amssymb,amsmath,floatfix,
groupedaddress,superscriptaddress]{revtex4-1}

\usepackage{amsmath}
\usepackage{graphicx}
\usepackage{xspace}

\begin{document}


\title{A 2-D  suspension of active agents: the role of fluid mediated  interactions}

\author{Hojjat Behmadi}
\affiliation{Physics Department, University of Zanjan, Zanjan 45371-38791, Iran.}

\author{Zahra Fazli}
\affiliation{Department of Physics, Institute for Advanced Studies in Basic Sciences (IASBS), Zanjan 45137-66731, Iran.}

\author{Ali Najafi}
\email[]{najafi@iasbs.ac.ir}
\affiliation{Department of Physics, Institute for Advanced Studies in Basic Sciences (IASBS), Zanjan 45137-66731, Iran.}
\affiliation{Physics Department, University of Zanjan, Zanjan 45371-38791, Iran.}
\date{\today}

\begin{abstract}
Taking into account simultaneously the  Vicsek short range ordering  and also the far-field  hydrodynamic  
interactions mediated by the ambient fluid, we investigate the role of long range interactions in the ordering phenomena in a quasi 2-dimensional active  suspension. 
By studying the number fluctuations, the velocity correlation functions and cluster size distribution function, 
we observe that depending on the number density of swimmers and the strength of noise, 
the hydrodynamic interactions affect the ordering in a suspension. For a fixed value of noise, at large density of particles, long range interactions enhance the clustering  in the system.  
\end{abstract}


%

\maketitle


\emph{Introduction.} Active agents that can convert a non-mechanical form of  energy to mechanical work, exhibit a wide range of interesting dynamical behaviors when they form a suspension \cite{Marchetti,Saintillan,Parrish1,moradi,peruani,nott}.  Both at the scales of macro and micro, there are many examples of such systems that have attracted enormous interests recently.  Schools of fishes and birds \cite{Weihs,Parrish2,Hall,Larkin}, bacterial suspensions \cite{Ben-Jacob,sano}, gels of 
cytoplasmic polymers \cite{frank-gel,kardar},  interacting  active Janus 
particles \cite{2Janus} and swimmers in non newtonian fluid \cite{ardekani1} 
are examples.  

One of the main questions that needs to be answered, is the nature of ordered phases  in such systems.  
In this article, we concentrate on the dynamics of micron-scale active agents.  
A wide class of works includes numerical simulations of micro-suspensions, based on phenomenological and simplified interaction terms between individual 
particles\cite{Ryan1,Hernandez-Ortiz1,Underhill,Ryan2,Ishikawa,Decoene1}. 
The well known model of Vicsek that
can correctly account the local ordering of elongated objects, built  the core of such studies\cite{Vicsek}. Such simulations reveal how a local ordering rule can 
lead the system to reach a state with large scale ordered phases \cite{Gregoire,Huepe,Chate1,Baglietto1,Baglietto2}.
 
Continuum  thermodynamic description of active suspensions, is  another line of approach that can address some macroscopic 
features of the systems\cite{Cisneros,Saintillan2,Hohenegger,Subramanian,Wensink,Drescher}. Dynamical equations derived   by  symmetry arguments or obtained from statistical averaging over microscopic forces, can capture the physics of ordered phases developed in such systems.
Only a few studies have investigated the role of long-range hydrodynamic interactions ({\rm HD}) between the particles \cite{baskaran2,sokolovaranson,golestan,holgerstark0}. 
Effects of such interactions is essential, specially in the case of micro-scale examples suspended in aqueous media. Fluid velocity 
produced by a moving particle, propagates instantaneously (a property of small scale hydrodynamic) through the medium and affects the motion of other particles. 
Some researchers, using a simple dipolar flow interaction mechanism, conclude that {\rm HD} may prevent the emergence of long-range order \cite{chateperuani}.  
The aim of this article is to improve our understanding of the ordering phenomena and collective behavior in a suspension of active microscopic agents.
To correctly account for such interactions, one needs to start from a hydrodynamic model that takes into account the internal structure of the swimmers.  
Starting from hydrodynamic interactions of a generic microscopic model, we numerically study the statistical parameters,
 that can reflect the nature of ordering in a suspension of such micron scale swimmers.
\begin{figure}
\centering
\includegraphics[width=.5\textwidth]{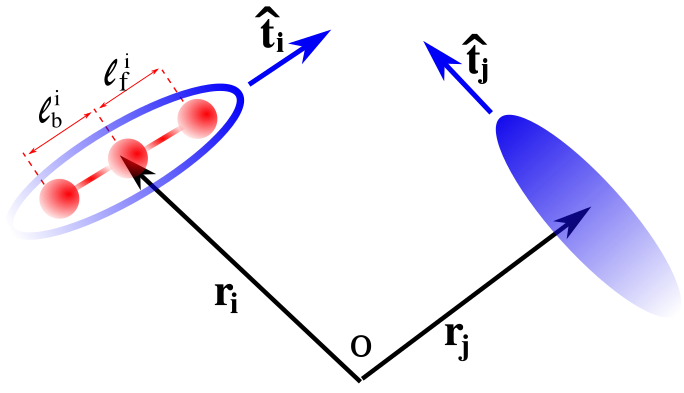}
\centering
\caption{The geometry and internal structure of two interacting swimmers are shown.
 Each swimmer, with two internal degrees of freedom, has an intrinsic swimming 
 direction denoted by $\hat{{\bf t}}$.
}
\label{fig1}
\end{figure}

\emph{Model.} 
To study the dynamics of a two dimensional collection of ${\cal N}$ interacting self propelled objects, we assume that the interaction between  
swimmers can be obtained from two-particle interactions. 
Fig.~\ref{fig1}, shows a schematic view of two swimmers that interact through both short and long range interactions. 
The position and the orientation of the $i$'th swimmer is shown by ${\bf r}_i$ and ${\hat {\bf t}}_i$, respectively. In addition to 
position and orientation, the internal structure of the swimmers is also important.
Each swimmer has an internal structure, that allows it to swim.
To model the internal structure of the swimmers, we use a 
minimal model with two internal degrees of freedom, namely the three beads connected by two arms\cite{2s,cicuta}. This is a generic model that can correctly explain the 
far field of both dipolar and quadrupolar  swimmers. Denoting the lengths of  front and back arms of a swimmer by 
$\ell_{f}^{i}(t)$ and $\ell_{b}^{i}(t)$ and the spheres radius by $a$, we can seek  internal motions that are able to propel the swimmer. As it is verified experimentally, a simple harmonic 
undulating motion  with a phase lag  on both arm lengths, is able to propel the swimmer at low Reynolds condition \cite{cicuta}. 
For identical swimmers, we choose an internal motion that is given by   $\ell_{f}^{i}(t)=\ell+u\sin(\omega t)$ and $\ell_{b}^{i}(t)=\ell(1+\delta)+u\sin(\omega t+\varphi_i)$, 
where   $\ell$ and $\ell(1+\delta)$ denote the average arm lengths, $u$ denotes the undulation amplitude, the frequency is shown by 
$\omega$ and the phase difference between the arms is denoted by $\varphi_i$. 
\begin{figure}[t]
\centering
\includegraphics[width=.5\textwidth]{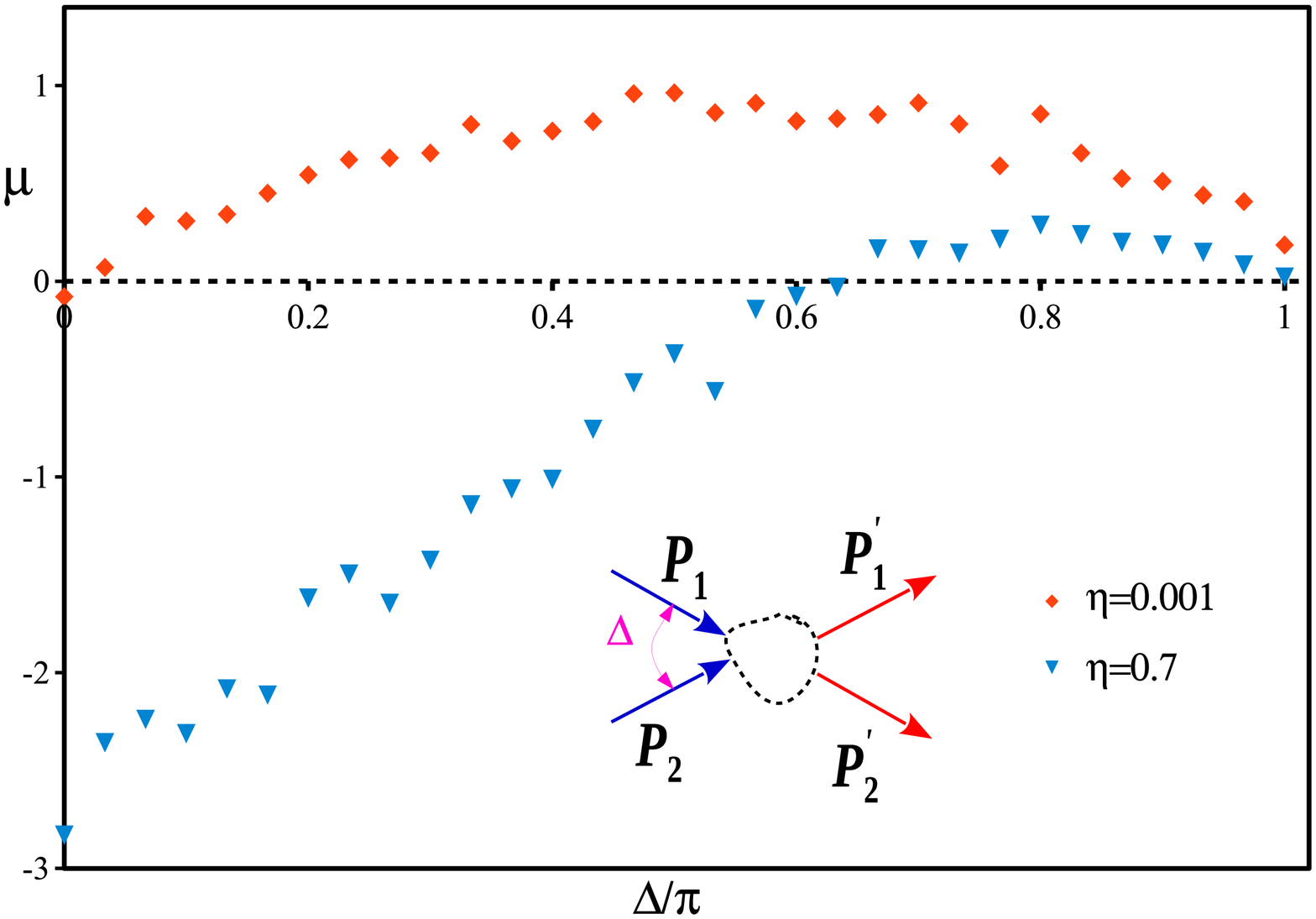}
\caption{Forward component of the change in total momentum, is plotted  as a function of incoming angular separation $\Delta$,  for 
different values of noise strength $\eta$.  At very low noise, the isotropic state is not stable. }
\label{fig2}
\end{figure}

The orientation of a swimmer in  a two dimensional reference frame can be represented by a single angle $\theta_i$. In this case we 
have: ${\hat {\bf t}}_i=(\cos\theta_i,~\sin\theta_i)$.
Detail hydrodynamic calculations (see appendix), show that the velocity of   $i$'th swimmer, moving in the presences of 
 $j$'th swimmer, can be written as \cite{farzin,yeomans}:
\begin{equation}
{\bf V}_i=v{\hat{\bf t}}_i+{\bf V}_{ij},~~~~\dot{\theta}_i=\Omega_{ij},\nonumber
\end{equation}
where the intrinsic swimming velocity of the swimmer depends on the internal structure: $v=v(a,u,\omega,\ell,\delta)$ and, the interaction 
terms are functions of the distance ${\bf r}_{ij}={\bf r}_{i}-{\bf r}_{j}$ and the orientation of swimmers: ${\bf V}_{ij}={\bf V}_{ij}(v,{\hat{\bf t}}_i,{\hat{\bf t}}_j,{\bf r}_{ij})$ and $\Omega_{ij}=\Omega_{ij}(v,{\hat{\bf t}}_i,{\hat{\bf t}}_j,{\bf r}_{ij})$ (See appendix for details). 
The interaction terms, ${\bf V}_{ij}$ and $\Omega_{ij}$, obtained with this kind of modeling are valid only for very far swimmers: $r_{ij}\gg \ell$.  Complexity of hydrodynamic equations, does not allow us to achieve 
analytical results for the short range part of the  interactions between swimmers. 

To overcome the complexity of short range 
hydrodynamic interactions, we approximate the short range part of the interactions 
by a very well known model of Vicsek interaction that is essentially a phenomenological 
short range interaction \cite{Vicsek}. 
This  interaction  enforces an elongated object (like what we have shown by ellipsoids in Fig.~\ref{fig1}) to 
change its direction  according to the average orientations of its neighbors.  
The Vicsek model does not fully consider all features of the short rang hydrodynamic interaction, but as an approximation we neglect other details of short range hydrodynamic interactions that are not included in Vicsek model. Vicsek's model 
mainly takes into account the steric interaction between the elongated objects. 
  To simplify our study, we assume that there is a crossover length $R_c$ that separate the short and long range forces. 
Two objects with separation smaller than this crossover length, 
interact with short range Vicsek model and beyond this length,  the long range hydrodynamic interactions are present.
Emergent collective motions of the Vicsek model are clearly known and 
our combined model here,  will show how  ${\rm HD}$ can affect such 
collective motions.
\begin{figure*}[t!]
\centering
        \includegraphics[width=.85\textwidth]{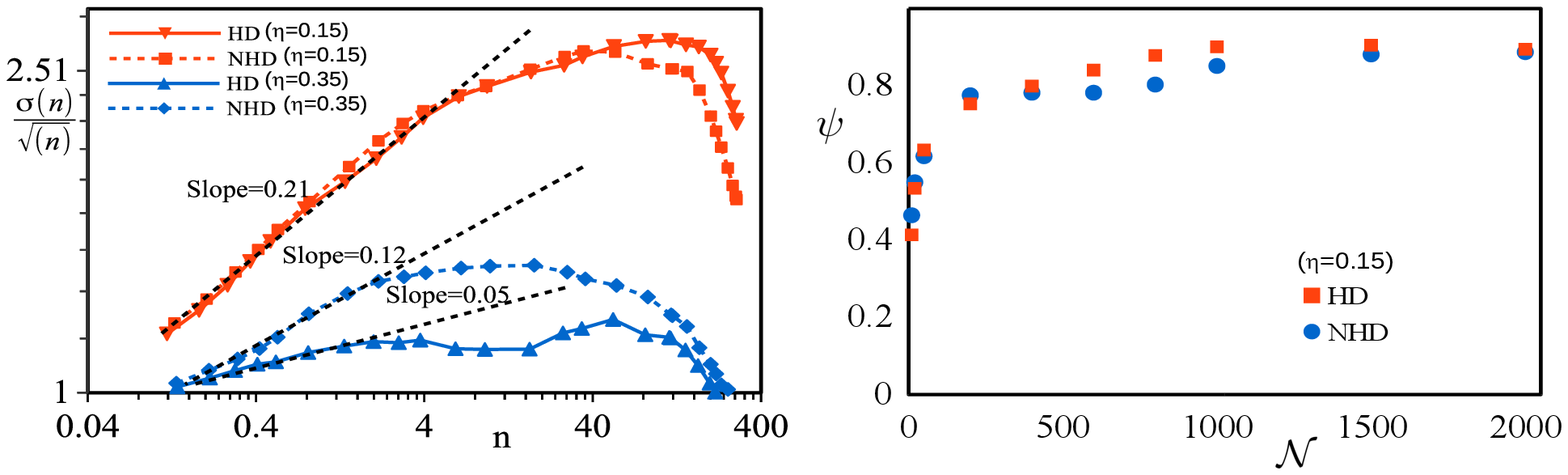}
    \caption{
Left:
    order parameter in terms of the number of swimmers.  
    Right:  density fluctuation as a function of average number is plotted  for
different strength of noise $\eta$.  Results are compared for two cases where the hydrodynamic
interaction is on or off ({\rm  HD} and {\rm NHD}).
}    
\label{fig3}
\end{figure*}
  
In order to numerically study the dynamics of a suspension, we  write the discrete dynamics of the $i$'th swimmer as: 
\begin{eqnarray} 
{\mathbf{r}}_{i}(t+\delta t)=\mathbf{r}_{i}(t)+ \delta t\left( 
v\mathbf{\hat{t}}_{i} + \sum_{j} \Theta(r_{ij}-R_c)\mathbf{V}_{ij}\right),\nonumber\\
{\theta}_{i}(t+\delta t)=\theta_{i}(t)+\theta_{i}^{V} + \eta \xi_{i}(t)+\delta t \sum_{j} \Theta(r_{ij}-R_c)\Omega _{ij},\nonumber
\end{eqnarray}
where Heaviside step function is $\Theta(x)=1$ for $x\ge 1$ and it is $0$ for $x<1$.
The short range contribution to the dynamics is given by: 
$\theta_{i}^{V}=\arg \sum'_{j}e^{i\theta_{j}(t)}$, 
where the summation runs  over all swimmers with $r_{ij}< R_c$.
We assume that the  fluctuations affect the 
dynamics of the swimmers, through a rotational noise represented by $\eta\xi_{i}(t)$. 
$\xi$ is random number with uniform probability in the interval $[-\pi,\pi]$ and the strength of noise $\eta$, can take any positive value.

\emph{Two swimmers scattering.} 
Before studying the case of many swimmers, it is instructive to start with two particles system.
Rich behavior emerging from long range hydrodynamic interaction, promises a non trivial behavior for the trajectories of two interacting swimmers.
A plethora of behavior, repulsive, attractive and oscillating trajectories can be observed. The details of such behavior has been studied 
extensively before \cite{farzin, yeomans}. 

An important feature that one can learn from the two body system, is the stability of isotropic phase in a many swimer system. In an isotropic phase, all swimmers moves in 
random directions and no direction is preferred.    
Using a kinetics theory approach with two body scatterings, it is shown that for a dilute system, the nature of two body scattering is the essential mechanism that determines the stability of the isotropic state \cite{Douchot}. 
Denoting by ${\bf P}$ and $\delta{\bf P}$, the initial total momentum and the change in total momentum after a binary scattering, we  
define  
the average forward component of the momentum change in a binary scattering by: $\mu=\langle{\bf P}\cdot \delta 
{\bf P}\rangle_{0}$.  Averaging is done over all impact parameters and as shown in 
fig.~\ref{fig2}(inset), the incoming angular separation is shown by $\Delta$.
Neglecting the self diffusion, for $\mu >0$, the isotropic state is unstable and the interactions will eventually lead the system to reach a 
polar state \cite{Douchot}.  For $\mu <0$, the interaction between particles is not able to develop a polar state.

Fig.~\ref{fig2}, shows the forward component of averaged change in momentum as a function of incoming angular separation $\Delta$. Here we have taken into account both 
the Vicsek and {\rm HD} interactions as described in the previous section. 
As one can see from this figure, for small noises, $\mu$ is positive and it reflects the instability of the isotropic state. Following such instability and for small noises, ordered state  (anisotropis phase) emerges. In anisotropic phase, 
the rotational symmetric is spontaneously broken and  all swimmers 
move in a preferred direction. 
Such an instability is a general feature of the Vicsek like interaction, and we see here that the long range hydrodynamic interaction does not affect the instability.
As one can see from the figure, by increasing the noise, $\mu$ starts to have negative values that reflects the stability 
of isotropic  state. This means that, still in the presence of {\rm HD}, we expect to 
observe ordered (anisotropic) phase at small values of noise. 

In the following parts we numerically investigate the detail role of {\rm HD} in the 
ordering of a suspension. 

\emph{Simulation.} 
To investigate the role of hydrodynamic interaction in the long time behavior of a  quasi 2-D suspension, we proceed by numerically simulating the system. Along this path we study  a set of order parameter and correlation functions. 
To quantify the polar order of the system, we define the polar order parameter as: 
$\psi=\frac{1}{{\cal N} v}|\sum_{i=1}^{\cal N}{\bf V}_i|$
fully polarized state (anisotropic phase) is given by $\psi=1$ and the isotropic state corresponds to $\psi=0$. 
Velocity autocorrelation and velocity-velocity correlation functions are defined as: 
\begin{eqnarray}
&&C_a(t)=\frac{1}{{\cal N}}\langle {\sum_{i=1}^{\cal N} \frac{{\bf V}_i(0)\cdot {\bf  V}_i(t)}{\vert\textbf{V}_i(0)\vert \, \vert\textbf{V}_i(t)\vert}} \rangle,\nonumber\\
&&C_{vv}(r)=\frac{1}{{\cal N}({\cal N}-1)}\langle {\sum_{i=1}^{\cal N} \sum_{j\neq i}^{\cal N}\frac{\textbf{V}_i(t)\cdot \textbf{V}_j(t)}{\vert\textbf{V}_i(t)\vert \, \vert\textbf{V}_j(t)\vert}} \rangle.\nonumber
\end{eqnarray}
where  $\langle\cdots\rangle$ denotes the averaging over all particles. 
These two correlation functions contain information about correlation time and 
correlation length in fluctuating system.
Local spatial ordering and clustering in the system can be understood in terms of 
the radial distribution function $g(r)$ that is defined by:
\begin{equation}
g(r)=\frac{\ell^2}{{\cal N}({\cal N}-1)}\langle {\sum_{i=1}^{\cal N} \sum_{j\neq i}^{\cal N} \delta(r-\vert \textbf{r}_{ij}\vert)} \rangle.\nonumber
\end{equation}
Defining all the required statistical parameters of our system, we will study the thermodynamic state of our system in the next section. 
 \begin{figure*}[t!]
 \centering
    \centering
\includegraphics[width=.85\textwidth]{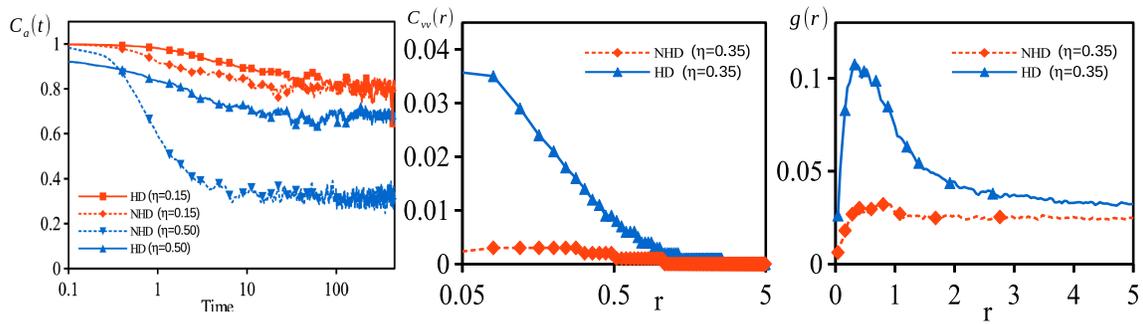}
    \caption{ Velocity auto-correlation function $C_a(t)$ (a), 
velocity-velocity correlation function $C_{vv}$ (b) and, radial distribution 
function $g(r)$ (c) are plotted     
 for a system with  
${\cal N}=2000$ and $\eta=0.35$. At large noises, both correlation time and 
correlation length and also the pairing strength are enhanced by {\rm HD}  interactions.      
  }
    \label{fig4}
\end{figure*}
In our numerical study, we consider a two dimensional suspension of   ${\cal N}$ particles in a square box of  
length  $L$ with periodic boundary condition has been considered. 
To make the equations non dimensional, we use $\ell$ and $v$ as characteristic length and velocity. 
In simulations, we choose a square box of size  $50\ell$ and  change the particle numbers from $100$ to $2000$. The time step in dimensionless units  is $\delta t = 0.001$ and a total number of $\sim 1.2\times 10^{6}$ steps is necessary to reach steady state.

\emph{Results.} 
Swarming behavior in our model, results from  interplay between hydrodynamic interactions, noise strength and number density 
of particles.
As our main goal here is to investigate the role of hydrodynamic interactions, we repeat all simulations with and without hydrodynamic interaction. In the first set of simulations, we perform the simulations  only with the Vicsek interaction then for the second set, 
we include the long range interactions as well. Comparison between  the results of these two sets of results will provide an understanding of the 
role of hydrodynamic interaction. All results marked by {\rm NHD} are obtained by taking into account the Vicsek interaction and the results marked by {\rm HD}, denote the cases where both Vicsek and long range hydrodynamic interactions are present.

Fig.~\ref{fig3}(a)  shows the polarization order parameter, $\psi$, as a function of  number of the swimmers (for fixed box size).
As we have expected from two-body scattering results have been obtained at previous section, for a fixed noise, increasing 
the density will result instability in the isotropic phase and a stable anisotropic 
polarized phase will appear. Our numerical results show that the appearance of such polarized state with $\psi\neq 0$, 
is not sensitive to long range part of the interaction. This is consistent with the 
results of previous section where, as we discussed, it is the short range part of the interaction 
that dominates in the instability mechanism for the isotropic phase.
The number fluctuation, is the other quantity that we have studied in our simulations. Denoting the average number of particles by $\langle n\rangle$, we study its  fluctuations: $\sigma=\langle n^2\rangle-\langle n\rangle^2$. 
This a statistical parameter that includes information about the non-equilibrium nature of a fluctuating system. The practical method which we have used to calculate  
the number fluctuations is as follows. For a given total number of swimmers ${\cal N}$, we start by a small window in the middle of the simulation box and measure both 
average number of particles and its fluctuation inside this window. Then changing the size of this window will allow us  to plot the number fluctuations as a function of average number. 
For a system that is in thermal equilibrium, we expect to see a relation like $\sigma\sim \langle n\rangle^\frac{1}{2}$. Fig.~\ref{fig3}(b) shows the results of numerical simulations, that as a result of particle's activity, deviate from equilibrium $\frac{1}{2}$ power law \cite{Ramaswamy}. 
The results, indicate that by  increasing the noise strength the system will tend to approach equilibrium power law (for larger noise, the slopes are smaller). 
 But for a large and  fixed noise strength, the {\rm HD} curve has a slope smaller than the {\rm NHD} curve. It turns out that for  this condition, the {\rm HD} interaction diminishes the out of equilibrium nature of the system.   This result critically depends on the strength of noise, our simulations shows 
that for smaller noise the {\rm HD} does not have any critical role in the behavior of number fluctuation.

Velocity auto-correlation function and  velocity-velocity correlation function, are plotted in 
fig.~\ref{fig4}(a) and (b). As one can see from the results, for large number density  
${\cal N}=2000$ and large noise $\eta=0.35$, the {\rm HD} interaction increases both the correlation time and correlation length. 
For small noise $\eta=0.15$ (results are given only for autocorrelation function), the increase in correlation time is very small.  
Fig.~\ref{fig4}(c), shows the results for radial distribution function $g(r)$. The strength of peak in this graph that corresponds to the 
pairing of the particles, strongly depends on the interaction between particles. {\rm HD} interaction increases the pairing and clustering in the system. 
\begin{figure*}[t!]
\centering
       \includegraphics[width=0.42\textwidth]{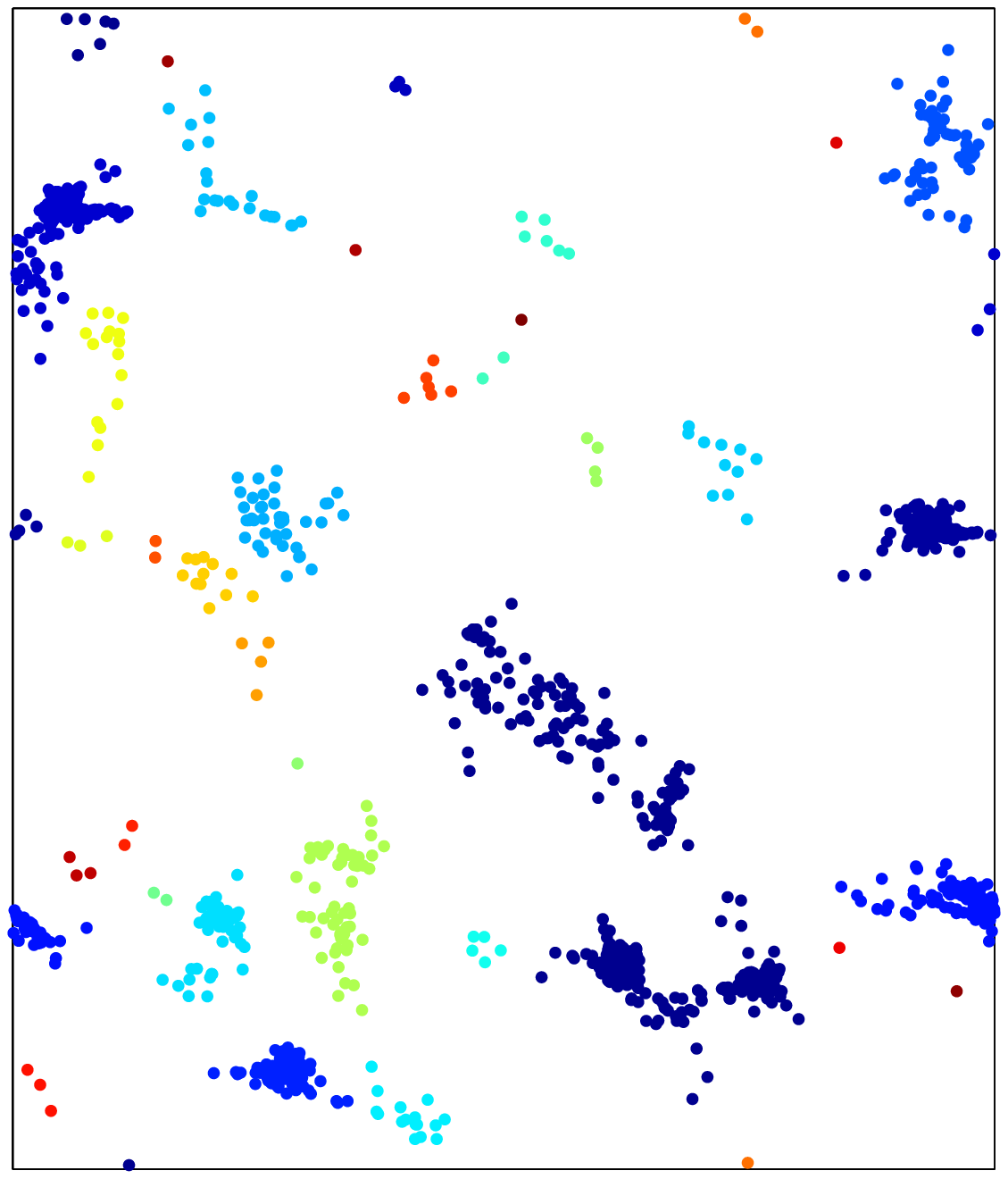}
              \includegraphics[width=0.42\textwidth]{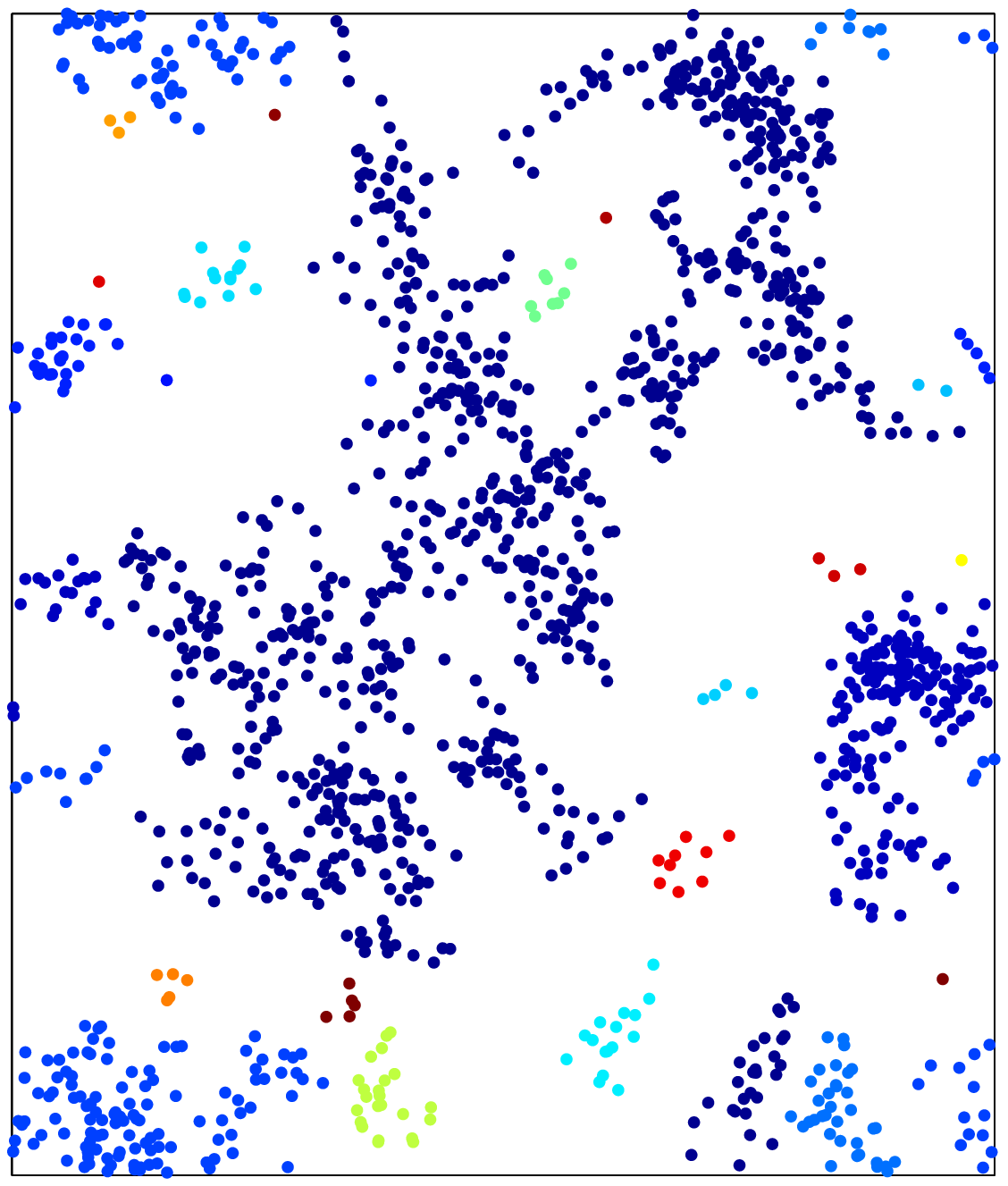}\\ \centering
       \includegraphics[width=0.85\textwidth]{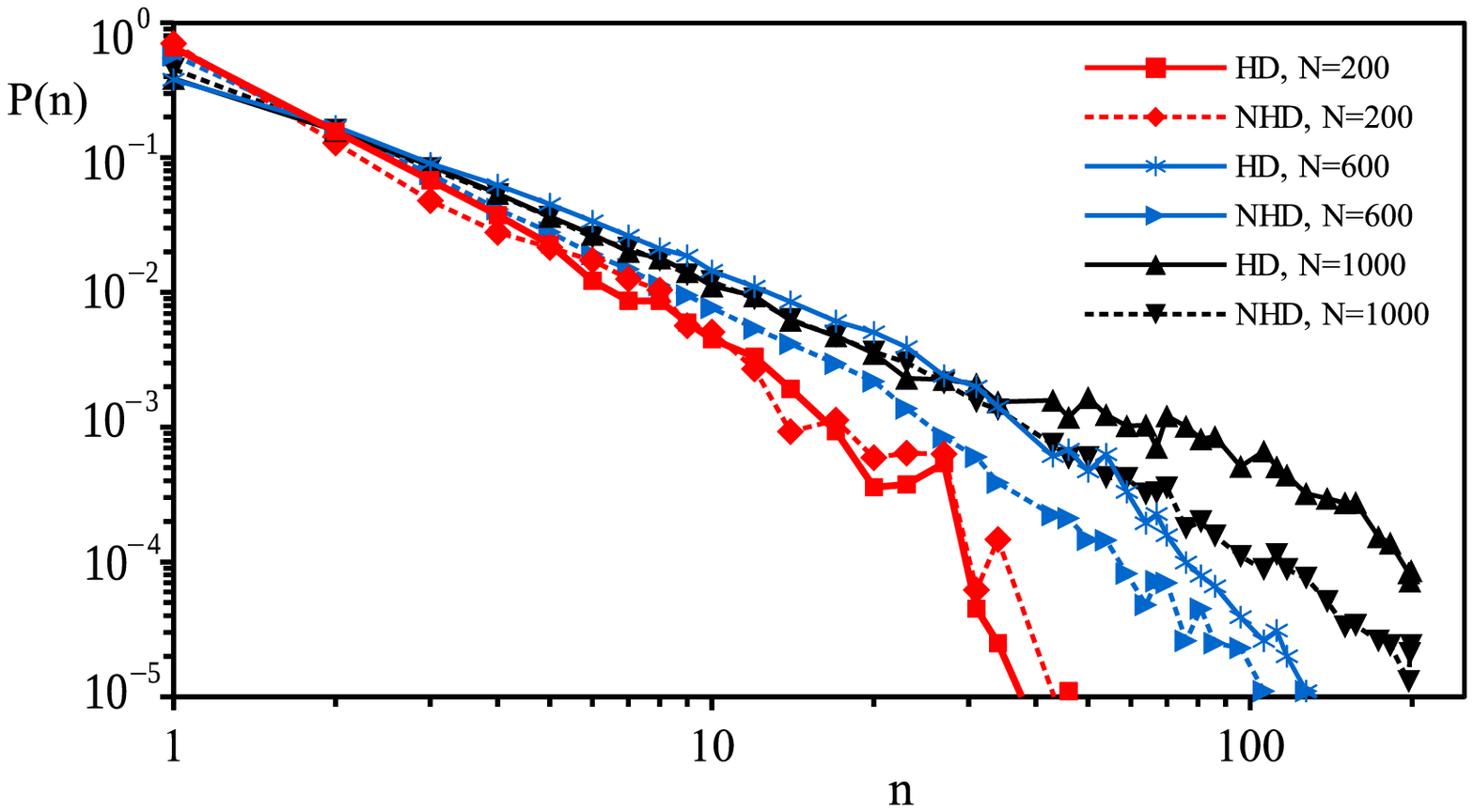}
\caption{Up: two snapshots of the system for ${\cal N}=1500$ with (left) and without (right) hydrodynamic interactions.  Different colors denote  different cluster and one can distinguish that {\rm HD} strongly enhances the clustering mechanism. Down: 
cluster size distribution function for various  particle number and noise strengths. For large number of particles, having large clusters are most frequent in the case of {\rm HD}.}
\label{fig5}
\end{figure*}

Size distribution function of clusters is
another important quantity that can help us in correctly analyzing the swarming behavior in a suspension. In a many body system of active agents, clusters of different sizes intermittently form and break. As a results of particle exchange between different clusters, a power law distribution function can be expected \cite{Huepe}. Intuitively, if two particles are within the alignment zone, they are considered to belong to a  same cluster. 
Size of a cluster $n$, is defined as the number of particles that belong to a same cluster.  
Examples of system snapshots, with and without {\rm HD} interactions are shown  fig.~\ref{fig5}(up), left and right, respectively. 
In  similar conditions, for the case where the {\rm HD} interaction is on, the clusters are finely distinguishable.  
Denoting by $P(n)$, the probability to have a cluster with size $n$, we study this function for different values of number density 
and  a fixed strength of noise $\eta=0.35$ in fig.~\ref{fig5}(down). Having clusters with large sizes, depends strongly on the interaction and the number density. For larger densities, {\rm HD} interaction increases the probability of finding large clusters, but by decreasing the density of particles, the {\rm HD} may change its role. Consistence with the previous result, at small density, the hydrodynamic interactions do not show any observable effects in our simulations. 

In conclusion, we have numerically studied the effects of long range hydrodynamic interactions in the ordering phenomena in an 
active suspension of micro-particles. We have shown that depending on the  strength of noise and number density of particles, the 
interactions have critical effects on the number fluctuations, correlation functions and clustering phenomena. Along this work, we are studying the effects of 
interplay between internal phases of the swimmers (here, we have assumed that all swimmer are in phase). Coherent effects observed in small systems \cite{coherentcoupling}, promise us to see 
interesting effects in suspensions.

\emph{Appendix.} 
Following the method given in references \cite{farzin,yeomans}, the intrinsic velocity of a single swimmer and it's hydrodynamic interaction with second swimmer can be written as:
$$
v=\frac{7}{24}\,(\frac{a}{\ell^2})\,(1+\delta)\,(u^2\,\omega)\,\sin\varphi_{i},
$$
\begin{eqnarray}
&&\mathbf{V}_{ij}=\alpha_1(\frac{\ell}{r_{ij}})^2\,\textbf{T}+{(\frac{\ell}{r_{ij}})}^3\,
 \left[ \alpha_2\,\textbf{D}+\alpha_3\,\textbf{E}\right],\nonumber\\
&&\boldsymbol\Omega_{ij}=\beta_1\,{(\frac{\ell}{r_{ij}})}^3\,\textbf{F}+{(\frac{\ell}{r_{ij}})}^4
 \left[\beta_2\,\textbf{G}+\beta_3\,\textbf{H}\right].\nonumber
\end{eqnarray}
The above results are obtained by  averaging over a complete period of  internal motion ($2\pi/\omega$).  In terms of  $\epsilon=\frac{a}{\ell^2}u^2\omega$ and $\varphi_{ij}=\varphi_i-\varphi_j$, the parameters are given by:
\begin{eqnarray}
&&\alpha_1=\frac{29}{64}\,(\frac{a\,\delta\epsilon}{\ell})\,\sin\varphi_j,~~~~~
\alpha_2=-\frac{\epsilon}{4}\,(\delta
+2)\,\sin\varphi_j,\nonumber\\
&&\alpha_3=\frac{\epsilon}{24}\,\left[(3+\delta)\,\sin\varphi_i+(3+2\delta)\,(\sin\varphi_j+\sin\varphi_{ij})\right],\nonumber\\
&&\beta_1=-\frac{29}{64}\,(\frac{a\,\delta\epsilon}{\ell^2})\,\sin\varphi_j,~~~~~\beta_2=\frac{3}{8}\,(\frac{\epsilon}{\ell})\,\sin\varphi_j\,(2-\delta),\nonumber\\
&&\beta_3=\frac{7}{48}\,(\frac{\epsilon}{\ell})\,\left[(3-\delta)\,\sin\varphi_i+(3-2\delta)(\sin\varphi_j+\sin\varphi_{ij})\right].\nonumber
\end{eqnarray}
and
\begin{equation}
\begin{array}{rcl}
&&\textbf{T}=-3 [M_{mn}\hat{t}_{jm}\hat{t}_{jn}] \,\hat{r}_{ij}\\\\
&
&\textbf{D}=-\frac{3}{2} [M_{mn} \hat{t}_{jm}\hat{t}_{jn}]\, \hat{t}_{j} +
\frac{3}{2} [M_{mnk} \hat{t}_{jm}\hat{t}_{jn}\hat{t}_{jk}]\, \hat{r}_{ij}\\\\
&
&\textbf{E}=-3 [M_{mn} \hat{t}_{jm}\hat{t}_{jn}]\, \hat{t}_i + 3 [M_{mnk}
\hat{t}_{jm}\hat{t}_{jn}\hat{t}_{ik}]\, \hat{r}_{ij}\\\\
&
&\textbf{F}=3 [M_{mnk}\, \hat{t}_{jm}\,\hat{t}_{jn}\,\hat{t}_{ik}]\,
(\hat{r}_{ij}\times\hat{t}_i),\\\\
&
&\textbf{G}= [M_{mnk} \hat{t}_{jm}\hat{t}_{jn}\hat{t}_{ik}]\,
(\hat{t}_j\times\hat{t}_i)-5 [M_{mnkl}
\hat{t}_{jm}\hat{t}_{jn}\hat{t}_{jk}\hat{t}_{il}]\,
(\hat{r}_{ij}\times\hat{t}_i),\\\\
&
&\textbf{H}=-\frac{15}{2} [M_{mnkl}
\hat{t}_{jm}\hat{t}_{jn}\hat{t}_{ik}\hat{t}_{il}]\, (\hat{r}_{ij}\times\hat{t}_i),\nonumber
\end{array}\nonumber
\end{equation}
where $\hat{t}_{i}$ represents the director of $i$'th swimmer and $\hat{t}_{ik}$ stands for its 
$k$'th component. In above relations, summation over indices $m,~n,~k,~l$ are assumed.
The symmetric and traceless tensors used at the above equations,  are given by:
\begin{equation}\label{tan}
\begin{array}{rcl}
&& \textit{M} _{ij}(\hat{r})=\hat{r}_i\hat{r}_j-\frac{1}{2}\delta_{ij},\\\\
&
&
\textit{M}_{ijk}(\hat{r})=4\hat{r}_i\hat{r}_j\hat{r}_k-(\delta_{jk}\hat{r}_i+\delta_{ik}\hat{r}_j+\delta_{ij}\hat{r}_k),\\\\
&
&
\textit{M}_{ijkl}(\hat{r})=6\hat{r}_i\hat{r}_j\hat{r}_k\hat{r}_l
+\frac{1}{4}(\delta_{ij}\delta_{kl}+\delta_{ik}\delta_{jl}+\delta_{il}\delta_{jk})\\
&
&
-(\delta_{ij}\hat{r}_k\hat{r}_l+\delta_{kl}\hat{r}_i\hat{r}_j+\delta_{ik}\hat{r}_j\hat{r}_l+\delta_{jl}\hat{r}_i\hat{r}_k+\delta_{jk}\hat{r}_i\hat{r}_l+\delta_{il}\hat{r}_j\hat{r}_k).\nonumber
\end{array}
\end{equation}
The above equations are valid for  two swimmers that are moving inside a 2-D plane. 
For numerical calculations we have set all 
phases to $\varphi_i=\pi/2$ and $\delta=0.1$.

\acknowledgments
We benefited from useful discussions with S. Ramaswamy and R. Golestanian. We also acknowledge Farideh Khalili and Mohsen Yarifard for their helps at the early stages of this work.


\bibliography{CollSwim} 
\end{document}